\documentclass[a4paper, 12pt, openany, oneside]{article}
\usepackage[cp1251]{inputenc} 
\usepackage[english, russian]{babel} 
\usepackage{amsmath, latexsym, amssymb, array, graphics,
amsfonts, amsmath, bm}

\makeatletter

\renewcommand{\Re}{\mathop{\rm Re\,}}

\newcommand{\Res}{\mathop{\rm Res\,}}
\makeatother
\usepackage{indentfirst}

\makeatother
\renewcommand{\baselinestretch}{1.2}
\usepackage[dvips]{graphicx}
\usepackage[flushleft,small]{caption2}

\begin{document}

 \thispagestyle{empty}
 \renewcommand{\abstractname}{\,}
\large

 \begin{center}
\bf The Second Stokes Problem with
Specular -- Diffusive Boundary Conditions in Kinetic Theory
\end{center}

\begin{center}
  \bf V. A. Akimova\footnote{$vikont\_ava@mail.ru$}
  A. V. Latyshev\footnote{$avlatyshev@mail.ru$} and
  A. A. Yushkanov\footnote{$yushkanov@inbox.ru$}
\end{center}\medskip

\begin{center}
{\it Faculty of Physics and Mathematics,\\ Moscow State Regional
University, 105005,\\ Moscow, Radio str., 10--A}
\end{center}\medskip

\tableofcontents
\setcounter{secnumdepth}{4}


\begin{abstract}

Получено решение полупространственной второй
задачи Стокса для одноатомного газа с
зеркально -- диффузными граничными условиями.
Вторая задача Стокса -- задача о поведении разреженного газа,
заполняющего полупространство. Плоскость,
ограничивающая полупространство, совершает гармонические колебания
в своей плоскости. Используется кинетическое уравнение с модельным
интегралом столкновений в форме $\tau$--модели.  Построена
функция распределения газовых молекул и найдена массовая скорость
газа в полупространстве.
Метод позволяет получить решение с произвольной степенью точности.
В основе метода лежит идея представления граничного условия на
функцию распределения в виде источника в кинетическом
уравнении. Решение получено в виде ряда Неймана.

{\it Ключевые слова:} вторая задача Стокса, зеркально - диффузные
граничные условия, ряды Неймана, функция распределения, скорость газа.


transport theory of gases.
\end{abstract}

\begin{center}
\item{}\section{Введение}
\end{center}

Задача о поведении газа над движущейся поверхностью в последние годы
привлекает пристальное внимание \cite{Stokes} -- \cite{15}. Это связано
с развитием современных технологий, в частности, технологий наноразмеров.
В \cite{Yakhot} -- \cite{15} эта задача решалась численными или
приближенными методами.

Впервые задача о поведении газа над стенкой, колеблющейся в своей плоскости,
была рассмотрена Дж. Г. Стоксом \cite{Stokes}. Задача решалась
гидродинамическим методом без учёта эффекта скольжения. Обычно
такую задачу называют второй задачей Стокса \cite{Yakhot}--\cite{SS-2002}.

В последние годы на тему этой задачи появился ряд публикаций.
В работе \cite{Yakhot} задача рассматривается для любых частот
колебания поверхности. Из кинетического уравнения БГК получено
уравнение типа гидродинамического. Рассматриваются гидродинамические
граничные условия. Вводится коэффициент, связывающий скорость газа на
поверхности со скоростью поверхности. Показано, что в случае высокочастотных
колебаний сила трения, действующая на поверхность, не зависит от частоты.

В работе \cite{SK-2007} получены коэффициенты вязкостного и теплового
скольжения с использованием различных модельных уравнений. Использованы
как максвелловские граничные условия, так и граничные условия
Черчиньяни --- Лэмпис.

В статье \cite{10} рассматривается газовый поток над бесконечной
пластиной, совершающей гармонические колебания в собственной плоскости.
Найдена скорость газа над поверхностью и сила, действующая
на поверхность со стороны газа. Для случая низких частот
задача решена на основе уравнения Навье --- Стокса.
Для произвольных скоростей колебаний поверхности задача решена
численными методами на основе кинетического уравнения Больцмана с
интегралом столкновений в форме БГК (Бхатнагар, Гросс, Крук).

Работа \cite{11} является экспериментальным исследованием.
Изучается поток газа, создаваемый механическим резонатором
при различных частотах колебания резонатора. Эксперименты
показывают, что при низких частотах колебаний резонатора,
действующая на него со стороны газа сила трения прямо пропорциональна
частоте колебания резонатора. При высоких частотах колебания
резонатора ($~10^8$ Гц) действующая на него сила трения от частоты
колебаний не зависит.

В последнее время задача о колебаниях плоской поверхности в
собственной плоскости изучается и для случая неньютоновских жидкостей \cite{5} и
\cite{6}.

В статье \cite{12} рассматривается пример практического применения
колебательной системы, подобной рассматриваемой во второй задаче
Стокса, в области нанотехнологий.

В диссертации \cite{15} были предложены два решения второй задачи
Стокса, учитывающие весь возможный диапазон коэффициента аккомодации
тангенциального импульса. Эти решения отвечают соответственно
гидродинамическому и кинетическому описанию поведения газа над
колеблющейся поверхностью в режиме со скольжением.

В наших работах \cite{ALY-1} и \cite{ALY-2} для второй задачи Стокса
отыскиваются собственные функции и соответствующие собственные
значения, отвечающие как дискретному, так и непрерывному спектрам.
Исследована структура дискретного и непрерывного спектров.
Развивается математический аппарат, необходимый для аналитического
решения задачи и приложений.

В настоящей работе строится аналитическое решение второй задачи Стокса.
На основе аналитического решения вычисляется скорость газа в
полупространстве и непосредственно у колеблющейся границы,
найдена сила трения, действующая со стороны газа на колеблющуюся пластину,
а также находится диссипация энергии пластины.

\begin{center}
\item{}\section{Постановка задачи}
\end{center}
\begin{center}
  \bf 1. Линеаризованное кинетическое уравнение для задачи о колебаниях газа
\end{center}
Задача о колебаниях газа решается в линеаризованной
постановке.
Линеаризация задачи проведена  при условии, что скорость газа много меньше
тепловой: $|u_y(t_1,x_1)|\ll v_T$,
где $v_T=1/\sqrt{\beta} $ -- тепловая скорость молекул $(\beta=m/(2kT))$,
имеющая порядок скорости звука.
Пусть разреженный одноатомный газ занимает полупространство $x>0$
над плоской твердой поверхностью, лежащей в плоскости $x=0$.
Поверхность $(y,z)$ совершает гармонические колебания вдоль оси $y$
по закону $u_s(t)=u_0e^{-i\omega t}$. Требуется построить функцию
распределения газовых молекул $f(t,x,\mathbf{v})$ и найти скорость газа
$u_y(t,x)$. Функция распределения ищется в виде
$f=f_0(1+\varphi)$, где $f_0$ -- абсолютный максвеллиан,
$f_0=n(\beta/\pi)^{3/2}\exp(-\beta v^2)$.

Рассмотрим линеаризованное кинетическое уравнение
$$
\dfrac{\partial \varphi}{\partial t}+
v_x\dfrac{\partial \varphi}{\partial x}+\nu\varphi(t,x,\mathbf{v})=
\dfrac{\nu m}{kT}v_yu_y(t,x),
\eqno{(1.1)}
$$
где $u_y(t,x)$ -- скорость газа,
$$
u_y(t,x)=\dfrac{1}{n}\int v_yf(t,x,\mathbf{v})d^3v,
$$
Здесь $\nu=1/\tau$ -- частота столкновений газовых молекул, $\tau$ --
время между двумя последовательными столкновениями молекул, $m$ --
масса молекулы, $k$ -- постоянная Больцмана, $T$ --
температура газа,
$n$ -- числовая плотность (концентрация) газа. Концентрация газа и
его температура считаются постоянными в линеаризованной постановке задачи.

Введем безразмерные скорости и параметры: безразмерную скорость молекул:
$\mathbf{C}=\sqrt{\beta}\mathbf{v}$ \;$(\beta=m/(2kT))$, безразмерную
массовую скорость $U_y(t,x)=\sqrt{\beta}u_y(t,x)$, безразмерное время
$t_1=\nu t$ и
безразмерную скорость колебаний пластины $U_s(t)=U_0e^{-i\omega t}$,
где $U_0=\sqrt{\beta}u_0$ -- безразмерная амплитуда скорости колебаний
границы полупространства. Тогда уравнение (1.1) может быть записано в виде:
$$
\dfrac{\partial \varphi}{\partial t_1}+
C_x\dfrac{\partial \varphi}{\partial x_1}+\varphi(t_1,x_1,\mathbf{C})=
{2C_y}U_y(t_1,x_1),
\eqno{(1.2)}
$$
где
$$
U_y(t_1,x_1)=\dfrac{1}{\pi^{3/2}}\int \exp(-C^2)C_y\varphi(t_1,x_1,
\mathbf{C})d^3C.
\eqno{(1.3)}
$$
Заметим, что для безразмерного времени $U_s(t_1)=U_0e^{-i \omega_1t_1}$.


С помощью (1.3) кинетическое линеаризованное уравнение (1.2) записывается в виде:
$$
\dfrac{\partial \varphi}{\partial t_1}+
C_x\dfrac{\partial \varphi}{\partial x_1}+\varphi(t_1,x_1,\mathbf{C}) =
\dfrac{2C_y}{\pi^{3/2}}
\int\exp(-{C'}^2)C_y'\varphi(t_1, x_1,\mathbf{C'})\,d^3C'.
\eqno{(1.4)}
$$

Сформулируем зеркально--диффузные граничные условия, записанные
относительно функции $\varphi(t_1,x_1,\mathbf{C})$:
$$
\varphi(t_1,0,\mathbf{C})=2qC_yU_s(t_1)+(1-q)\varphi(t_1,0,
-C_x,C_y,C_z),\quad C_x>0,
\eqno{(1.5)}
$$
и
$$
\varphi(t_1,x_1\to+\infty,\mathbf{C})=0.
\eqno{(1.6)}
$$

Итак, граничная задача о колебаниях газа сформулирована полностью и
состоит в решении уравнения (1.4) с граничными условиями (1.5) и (1.6).

\begin{center}
\bf 2. Декомпозиция граничной задачи
\end{center}

Учитывая, что колебания пластины рассматриваются вдоль оси $y$,
будем искать, следуя Черчиньяни \cite{16},
функцию $\varphi(t_1,x_1,\mathbf{C})$ в виде
$$
\varphi(t_1,x_1,\mathbf{C})=C_yH(t_1,x_1,C_x).
\eqno{(2.1)}
$$
Тогда безразмерная скорость газа (1.3) с помощью (2.1) равна
$$
U_y(t_1,x_1)=\dfrac{1}{2\sqrt{\pi}}\int\limits_{-\infty}^{\infty}
\exp(-C_x'^2)H(t_1,x_1,C_x')dC_x'.
\eqno{(2.2)}
$$

С помощью указанной выше подстановки (2.1) кинетическое уравнение (1.4)
преобразуется к виду:
$$
\dfrac{\partial H}{\partial t_1}+C_x\dfrac{\partial H}{\partial x_1}+
H(t_1,x_1,C_x)=\dfrac{1}{\sqrt{\pi}}\int\limits_{-\infty}^{\infty}
\exp(-C_x'^2)H(t_1,x_1,C_x')dC_x'.
\eqno{(2.3)}
$$

Граничные условия (1.5) и (1.6) преобразуются в следующие:
$$
H(t_1,0,C_x)=2qU_s(t_1)+(1-q)H(t_1,0,-C_x),\qquad C_x>0,
\eqno{(2.4)}
$$
$$
H(t_1,x_1\to +\infty, C_x)=0.
\eqno{(2.5)}
$$

Следующим шагом выделим временную переменную, положив далее:
$$
H(t_1,x_1,C_x)=e^{-i\omega_1t_1}h(x_1,C_x).
\eqno{(2.6)}
$$

Теперь мы получаем уравнение относительно функции $h(x_1,C_x)$:
$$
C_x\dfrac{\partial h}{\partial x_1}+z_0h(x_1,C_x)
=\dfrac{1}{\sqrt{\pi}}\int\limits_{-\infty}^{\infty}
\exp(-C_x'^2)h(x_1,C_x')dC_x',
\eqno{(2.7)}
$$
где
$$
z_0=1-i\omega_1.
$$

Граничные условия (2.4) и (2.5) переходят в следующие:
$$
h(0,C_x)=2qU_0+(1-q)h(0,-C_x),\qquad C_x>0,
\eqno{(2.8)}
$$
и
$$
h(x_1\to+\infty,C_x)=0.
\eqno{(2.9)}
$$

Тогда безразмерная массовая скорость равна:
$$
U_y(t_1,x_1)=e^{-i\omega_1t_1}U(x_1),
$$
где
$$
U(x_1)=\dfrac{1}{2\sqrt{\pi}}\int\limits_{-\infty}^{\infty}
\exp(-C_x'^2)h(x_1,C_x')dC_x'.
\eqno{(2.10)}
$$

Мы получили граничную задачу, состоящую в решении уравенния (2.7)
с граничными условиями (2.8) и (2.9).
Далее безразмерную координату $x_1$ снова будем обозначать через
$x$.
Продолжим функцию $h(x,\mu)$ на "отрицательное"\, полупространство, полагая
$$
h(x,\mu)=h(-x,-\mu), \qquad \mu>0.
$$

Сформулируем зеркально -- диффузные граничные условия для функции
распределения соответственно для "положительного"\, ($x>0$) и для
"отрицательного"\,($x<0$) полупространств:
$$
h(+0,\mu)=2qU_0+(1-q)h(+0,-\mu)=2qU_0+(1-q)h(-0,\mu), \quad \mu>0,
\eqno{(2.11)}
$$
$$
h(-0,\mu)=2qU_0+(1-q)h(-0,-\mu)=2qU_0+(1-q)h(+0,\mu), \quad \mu<0.
\eqno{(2.12)}
$$

Мы получили две граничные задачи: одна задача описывается уравнением
(2.7) с граничными условиями (2.9) и (2.11), вторая описывается уравнением
(2.7) с граничными условиями (2.9) и (2.12). Первая задача определена в
полупространстве $x>0$, вторая -- в полупространстве $x<0$. Эти две задачи
сформулируем в виде одной задачи. Для этого
включим граничные условия (2.11) и (2.12) в кинетическое
уравнение следующим образом:
$$
\mu \dfrac{\partial h}{\partial x}+z_0h(x,\mu)=2U(x)+
|\mu|\Big[2qU_0-q h(\mp 0,\mu)\Big]
\delta(x),
\eqno{(2.13)}
$$
где $\delta(x)$ -- дельта--функция Дирака. Верхний знак минус в (2.13)
отвечает первой задаче в полупространстве $x>0$, а нижний знак плюс --
второй задаче в полупространстве $x<0$.
\begin{center}
\item{}\section{Кинетическое уравнение во втором и четвертом
квадрантах фазового пространства}
\end{center}

Решая уравнение (2.13) при $x>0,\,\mu<0$, считая заданным массовую
скорость $U(x)$, получаем, удовлетворяя граничным условиям,
следующее решение:
$$
h^+(x,\mu)=-\dfrac{1}{\mu}\exp(-\dfrac{z_0x}{\mu})
\int\limits_{x}^{+\infty} \exp(+\dfrac{z_0t}{\mu})2U(t)\,dt.
\eqno{(3.1)}
$$

Аналогично при $x<0,\,\mu>0$ находим:
$$
h^-(x,\mu)=\dfrac{1}{\mu}\exp(-\dfrac{z_0x}{\mu})
\int\limits_{-\infty}^{x} \exp(+\dfrac{z_0t}{\mu})2U(t)\,dt.
\eqno{(3.2)}
$$

Теперь уравнение (2.13)  можно переписать, заменив второй член в
квадратной скобке из (2.13) согласно (3.1) и (3.2), в виде:
$$
\mu\dfrac{\partial h}{\partial x}+z_0h(x,\mu)=2U(x)+
|\mu|\Big[2qU_0-qh^{\pm}(0,\mu)\Big]\delta(x).
\eqno{(3.3)}
$$

В уравнении (3.3) граничные значения $h^{\pm}(0,\mu)$
выражаются через массовую скорость:
$$
h^{\pm}(0,\mu)=-\dfrac{1}{\mu}e^{-z_0x/\mu} \int\limits_{0}^{\pm
\infty}e^{z_0t/\mu}2U(t)dt=h(\pm 0,\mu).
$$

Решение уравнения (3.3) ищем в виде интегралов Фурье:
$$
2U(x)=\dfrac{1}{2\pi}\int\limits_{-\infty}^{\infty}
e^{ikx}E(k)\,dk,\qquad
\delta(x)=\dfrac{1}{2\pi}\int\limits_{-\infty}^{\infty}
e^{ikx}\,dk,
\eqno{(3.4)}
$$
$$
h(x,\mu)=\dfrac{1}{2\pi}\int\limits_{-\infty}^{\infty}
e^{ikx}\Phi(k,\mu)\,dk.
\eqno{(3.5)}
$$

При этом функция распределения $h^+(x,\mu)$ выражается через
спектральную плотность $E(k)$ массовой скорости следующим образом:
$$
h^+(x,\mu)=-\dfrac{1}{\mu}\exp(-z_0\dfrac{x}{\mu})
\int\limits_{x}^{+\infty} \exp(+\dfrac{z_0t}{\mu})dt
\dfrac{1}{2\pi}
\int\limits_{-\infty}^{+\infty}e^{ikt}E(k,\mu)\,dk=
$$
$$
=\dfrac{1}{2\pi}\int\limits_{-\infty}^{\infty}\dfrac{e^{ikx}
E(k,\mu)}{z_0+ik\mu}dk.
$$

Аналогично,

$$
h^-(x,\mu)=\dfrac{1}{2\pi}
\int\limits_{-\infty}^{\infty}\dfrac{e^{ikx}
E(k,\mu)}{z_0+ik\mu}dk.
$$

Таким образом,
$$
h^{\pm}(x,\mu)=\dfrac{1}{2\pi}
\int\limits_{-\infty}^{\infty}\dfrac{e^{ikx}
E(k,\mu)}{z_0+ik\mu}dk.
\eqno{(3.6)}
$$
\begin{center}
\item{}\section{Характеристическая система}
\end{center}
Теперь подставим интегралы Фурье (3.5) и (3.4), а также равенство
(3.6) в уравнение (3.3) и определение функции $U(x)$ (см. (2.10)).
Получаем характеристическую систему уравнений:
$$
\Phi(k,\mu)(z_0+ik\mu)=
E(k)+|\mu|\Big[2qU_0
-\dfrac{q}{2\pi}
\int\limits_{-\infty}^{\infty}\dfrac{E(k_1)dk_1}{z_0+ik_1\mu}\Big],
\eqno{(3.1)}
$$
$$
E(k)=\dfrac{1}{\sqrt{\pi}}\int\limits_{-\infty}^{\infty}
e^{-t^2}\Phi(k,t)dt.
\eqno{(3.2)}
$$

Из уравнения (3.1) выразим $\Phi(k,\mu)$ и подставим в (3.2). Получаем
характеристическое уравнение:
$$
E(k)L(k)=2qU_0T_1(k)
-\dfrac{q}{2\pi^{3/2}}\int\limits_{-\infty}^{\infty}
E(k_1)dk_1\int\limits_{-\infty}^{\infty}
\dfrac{e^{-t^2}|t|dt}{(z_0+ikt)(z_0+k_1t)}.
\eqno{(3.3)}
$$
Здесь
$$
T_n(k)=\dfrac{1}{\sqrt{\pi}}\int\limits_{-\infty}^{\infty}
\dfrac{e^{-t^2}t^n\,dt}{z_0+ikt},\quad n=1,2,\cdots,\quad
L(k)=1-\dfrac{1}{\sqrt{\pi}}\int\limits_{-\infty}^{\infty}
\dfrac{e^{-t^2}dt}{z_0+ikt}.
$$
Нетрудно видеть, что
$$
L(k)=
\dfrac{1}{\sqrt{\pi}}\int\limits_{-\infty}^{\infty}e^{-t^2}dt-
\dfrac{1}{\sqrt{\pi}}\int\limits_{-\infty}^{\infty}\dfrac{e^{-t^2}dt}
{z_0+ikt}=\dfrac{1}{\sqrt{\pi}}\int\limits_{-\infty}^{\infty}
\dfrac{e^{-t^2}i(kt-\omega)\;dt}{z_0+ikt}.
$$

Кроме того, внутренний интеграл в (3.3) обозначим:
$$
J(k,k_1)=\dfrac{1}{\sqrt{\pi}}\int\limits_{-\infty}^{\infty}
\dfrac{e^{-t^2}|t|dt}{(z_0+ikt)(z_0+k_1t)}.
$$

Заметим, что $J(k,0)=T_1(k), \; J(0,k_1)=T_1(k_1)$.
Перепишем теперь уравнение (3.3) с помощью предыдущего равенства в
следующем виде:
$$
E(k)L(k)+\dfrac{q}{2\pi}
\int\limits_{-\infty}^{\infty} J(k,k_1)E(k_1)\,dk_1=2qU_0T_1(k).
\eqno{(3.4)}
$$

Уравнение (3.4) -- интегральное уравнение Фредгольма второго
рода.

Укажем на связь между функцией $L(k)$ и дисперсионной функцией
$\lambda(z)$ \cite{ALY-1}--\cite{ALY-3}. Представим функцию $L(k)$ в виде:
$$
L(k)=1-\dfrac{1}{\sqrt{\pi}ik}\int\limits_{-\infty}^{\infty}
\dfrac{e^{-t^2}dt}{t-z},\quad\text{где}\quad
z=-\dfrac{1-i\omega_1}{ik}.
$$
Тогда
$$
L(k)=1+\dfrac{z}{\sqrt{\pi}(1-i\omega_1)}\int\limits_{-\infty}^{\infty}
\dfrac{e^{-t^2}dt}{t-z}=\dfrac{\lambda(z)}{z_0}.
$$

\begin{center}
\item{}\section{Ряды Неймана}
\end{center}

Разложим решение характеристической системы уравнений (3.4) и (3.1) в ряды
по степеням коэффициента диффузности $q$:
$$
E(k)=2U_0q[E_0(k)+q\,E_1(k)+q^2\,E_2(k)+\cdots],
\eqno{(4.1)}
$$
$$
\Phi(k,\mu)=2U_0q[\Phi_0(k,\mu)+q\,\Phi_1(k,\mu)+q^2\,\Phi_2(k,\mu)+\cdots].
\eqno{(4.2)}
$$

Подставим ряды (4.1) и (4.2) в уравнения (3.4) и (3.1). Получаем, что
$$
\Big[E_0(k)+qE_1(k)+q^2E_2(k)+\cdots+ \Big]L(k)=$$$$=T_1(k)-\dfrac{q}{2\pi}
\int\limits_{-\infty}^{\infty}J(k,k_1)\Big[E_0(k_1)+qE_1(k_1)+q^2E_2(k_1)+
\cdots+\Big]dk_1,
\eqno{(4.3)}
$$
$$
\Big[\Phi_0(k,\mu)+q\,\Phi_1(k,\mu)+q^2\,\Phi_2(k,\mu)+\cdots\Big](z_0+ik\mu)=
$$$$=\Big[E_0(k)+qE_1(k)+q^2E_2(k)+\cdots\Big]+
$$
$$+
|\mu|-|\mu|q\dfrac{1}{2\pi}\int\limits_{-\infty}^{\infty}
\dfrac{E_0(k_1)+qE_1(k_1)+q^2E_2(k_1)+\cdots}{z_0+ik_1\mu}dk_1.
\eqno{(4.4)}
$$

Приравнивая коэффициенты в левой и правой частях уравнений (4.3) и (4.4)
при одинаковых степенях $q$, получаем систему зацепленных уравнений. При
$q^0$ получаем:
$$
E_0(k)L(k)=T_1(k),
\eqno{(4.5)}
$$
$$
\Phi_0(k,\mu)(z_0+ik\mu)=E_0(k)+|\mu|.
\eqno{(4.6)}
$$
При $q$ получаем:
$$
E_1(k)L(k)=-\dfrac{1}{2\pi}\int\limits_{-\infty}^{\infty}J(k,k_1)E_0(k_1)dk_1,
\eqno{(4.7)}
$$
$$
\Phi_1(k,\mu)(z_0+ik\mu)=E_1(k)-|\mu|\dfrac{1}{2\pi}
\int\limits_{-\infty}^{\infty}\dfrac{E_0(k_1)dk_1}{z_0+ik_1\mu},
\eqno{(4.8)}
$$
При $q^2$ получаем:
$$
E_2(k)L(k)=-\dfrac{1}{2\pi}\int\limits_{-\infty}^{\infty}J(k,k_1)E_1(k_1)dk_1,
\cdots,
\eqno{(4.9)}
$$
$$
\Phi_2(k,\mu)(z_0+ik\mu)=E_2(k)-|\mu|\dfrac{1}{2\pi}
\int\limits_{-\infty}^{\infty}\dfrac{E_1(k_1)dk_1}{z_0+ik_1\mu},\cdots
\eqno{(4.10)}
$$
При $q^n$ получаем:
$$
E_n(k)L(k)=
-\dfrac{1}{2\pi}\int\limits_{-\infty}^{\infty}J(k,k_1)E_{n-1}(k_1)dk_1,
\eqno{(4.11)}
$$
$$
\Phi_n(k,\mu)(z_0+ik\mu)=E_n(k)-|\mu|\dfrac{1}{2\pi}
\int\limits_{-\infty}^{\infty}\dfrac{E_{n-1}(k_1)dk_1}{z_0+ik_1\mu},\cdots.
\eqno{(4.12)}
$$

Массовую скорость $U(x)$ и функцию распределения также разложим в ряды по
степеням $q$:
$$
U(x)=qU_0\Big[U_0(x)+qU_1(x)+q^2U_2(x)+\cdots\Big],
\eqno{(4.13)}
$$
$$
h(x,\mu)=qU_0\Big[h_0(x,\mu)+qh_1(x,\mu)+q^2h_2(x,\mu)+\cdots\Big].
\eqno{(4.14)}
$$

\begin{center}
  \item{}\section{Построение рядов Неймана}
\end{center}

\begin{center}
  \item{}\subsection{Нулевое приближение}
\end{center}

Из формулы (4.2) для нулевого приближения находим:
$$
E_0(k)=\dfrac{T_1(k)}{L(k)}.
\eqno{(5.1)}
$$

Нулевое приближение массовой скорости на основании (4.1), (3.4) и (4.13) равно:
$$
U^{(0)}(x)=qU_0U_0(x)=qU_0\dfrac{1}{2\pi}\int\limits_{-\infty}^{\infty}
e^{ikx}E_0(k)\,dk=qU_0\dfrac{1}{2\pi}
\int\limits_{-\infty}^{\infty}
e^{ikx}\dfrac{T_1(k)}{L(k)}dk.
\eqno{(5.2)}
$$

Согласно (4.6) находим:
$$
\Phi_0(k,\mu)= \dfrac{E_0(k)+|\mu|}
{z_0+ik\mu},
$$
и, следовательно, нулевое приближение функции распределения $h(x,\mu)$
на основании (3.5) и (4.14) равно:
$$
h^{(0)}(x,\mu)=qU_0h_0(x,\mu)=\dfrac{qU_0}{\pi}\int\limits_{-\infty}^{\infty}
\dfrac{E_0(k)+|\mu|}{z_0+ik\mu}e^{ikx}dk.
$$

\begin{center}
  \item{}\subsection{Первое приближение}
\end{center}

Перейдем к первому приближению. В первом приближении из уравнения
(4.7) находим:
$$
E_1(k)=-\dfrac{1}{2\pi L(k)}\int\limits_{-\infty}^{\infty}
\dfrac{J(k,k_1)}{L(k_1)}T_1(k_1)dk_1,
$$
а из уравнения (4.8) получаем:
$$
\Phi_1(k,\mu)=\dfrac{1}{z_0+ik\mu}\Bigg[E_1(k)-\dfrac{|\mu|}{2\pi}
\int\limits_{-\infty}^{\infty}\dfrac{E_0(k_1)dk_1}{z_0+ik_1\mu}\Bigg].
$$

Первая поправка к нулевому приближению массовой скорости равна:
$$
U_1(x)=\dfrac{1}{2\pi}\int\limits_{-\infty}^{\infty}e^{ikx}E_1(k)dk=
-\dfrac{1}{(2\pi)^2}\int\limits_{-\infty}^{\infty}\dfrac{e^{ikx}}{L(k)}
\int\limits_{-\infty}^{\infty}J(k,k_1)\dfrac{T_1(k_1)}{L(k_1)}dk_1,
$$
а к функции распределения такова:
$$
h_1(x,\mu)=\dfrac{1}{\pi}\int\limits_{-\infty}^{\infty}e^{ikx}\Phi_1(k,\mu)dk=
\dfrac{1}{\pi}\int\limits_{-\infty}^{\infty}\dfrac{e^{ikx}}{z_0+ik\mu}\times
$$
$$
\times\Bigg[-\dfrac{1}{2\pi L(k)}
\int\limits_{-\infty}^{\infty}J(k,k_1)E_0(k_1)dk_1
-\dfrac{|\mu|}{2\pi}\int\limits_{-\infty}^{\infty}\dfrac{E_0(k_1)dk_1}
{z_0+ik_1\mu}\Bigg].
$$

Итак, в линейном приближении массовая скорость вычисляется по формуле
$$
U(x)=qU_0[U_0(x)+qU_1(x)]=qU_0\Big[\dfrac{1}{2\pi}
\int\limits_{-\infty}^{\infty}e^{ikx}E_0(k)dk+\dfrac{q}{2\pi}
\int\limits_{-\infty}^{\infty}e^{ikx}E_1(k)dk\Big],
$$
а функция распределения -- по формуле
$$
h(x,\mu)=qU_0\Bigg[\dfrac{1}{\pi}\int\limits_{-\infty}^{\infty}
e^{ikx}\Phi_0(k,\mu)dk+\dfrac{q}{\pi}\int\limits_{-\infty}^{\infty}
\Phi_1(k,\mu)dk\Bigg].
$$

\begin{center}
  \item{}\subsection{Второе и высшие приближения}
\end{center}

Из уравнений (4.9) и (4.10) получаем:
$$
E_2(k)=-\dfrac{1}{2\pi L(k)}\int\limits_{-\infty}^{\infty}J(k,k_1)E_1(k_1)dk_1=
$$
$$
=\dfrac{1}{(2\pi)^2}\int\limits_{-\infty}^{\infty}\int\limits_{-\infty}^{\infty}
\dfrac{J(k,k_1)J(k_1,k_2)}{L(k_1)L(k_2)}T_1(k_2)dk_1dk_2,
$$
$$
\Phi_2(k,\mu)=\dfrac{1}{z_0+ik\mu}\Bigg[E_2(k)-\dfrac{|\mu|}{2\pi}
\int\limits_{-\infty}^{\infty}\dfrac{E_1(k_1)dk_1}{z_0+ik_1\mu}\Bigg].
$$

Из уравнений (4.11) и (4.12) получаем:
$$
E_n(k)=\dfrac{(-1)^n}{(2\pi)^nL(k)}\int\limits_{-\infty}^{\infty}\cdots
\int\limits_{-\infty}^{\infty}\dfrac{J(k,k_1)J(k_1,k_2)\cdots
J(k_{n-1},k_n)}{L(k_1)L(k_2)\cdots L(k_n)}T_1(k_n)dk_1\cdots dk_n
$$
и
$$
\Phi_n(k,\mu)=\dfrac{1}{z_0+ik\mu}\Bigg[E_n(k)-\dfrac{|\mu|}{2\pi}
\int\limits_{-\infty}^{\infty}\dfrac{E_{n-1}(k_1)dk_1}{z_0+ik_1\mu}\Bigg].
$$

На основании двух последних формул заключаем, что ряды (4.13) и (4.14)
для массовой скорости и функции распределения соответственно построены.

\begin{center}
  \item{}\subsection{Анализ решения в предельном случае больших частот}
\end{center}

Рассмотрим полученное решение второй задачи Стокса в случае больших частот
колебаний ограничивающей газ пластины: $\omega_1\gg 1$; тогда $z_0\approx -i
\omega_1$. В этом случае $L(k)\approx 1$. Поэтому, меняя порядок
интегрирования в формуле (5.2), имеем:
$$
\dfrac{U^{(0)}(x)}{U_0}=\dfrac{q}{2\pi}\int\limits_{-\infty}^{\infty}
e^{ikx}dk\dfrac{1}{\sqrt{\pi}}\int\limits_{-\infty}^{\infty}
\dfrac{e^{-t^2}tdt}{z_0+ikt}=
$$
$$
=\dfrac{q}{2\pi\sqrt{\pi}}\int\limits_{-\infty}^{\infty}e^{-t^2}tdt
\int\limits_{-\infty}^{\infty}\dfrac{e^{ikx}dk}{z_0+ikt}.
$$
Вычислим внутренний интеграл:
$$
\dfrac{1}{2\pi i}\int\limits_{-\infty}^{\infty}\dfrac{e^{ikx}dk}{z_0+ikt}=
\Res_{k=iz_0/t}\dfrac{e^{ikx}}{z_0+ikt}=\left\{
\begin{array}{c}
  \dfrac{e^{-xz_0/t}}{it},\quad t > 0, \\
  0,\quad t<0
\end{array}\right.
$$
Следовательно, в нулевом приближении для  получаем:
$$
\dfrac{U^{(0)}(x)}{U_0}=\dfrac{q}{\sqrt{\pi}}\int\limits_{0}^{\infty}
\exp\Big(-t^2+i\dfrac{x\omega_1}{t}\Big)dt.
\eqno{(5.3)}
$$

Из формулы (5.3) следует, что для  скорости газа на поверхности
колеблющейся плоскости справедлива формула:
$$
u_y(0)=q\dfrac{u_0}{2}.
$$

Скорость газа представим в виде $u_y(x)=u_0w_y(x)$, где
$w_y(x)$ -- нормированная на амплитуду $u_0$ скорость газа в полупространстве
$x>0$. 
На рис. 1 и 2 изобразим зависимость нормированной  скорости газа $w_y(x)$: 
$$
w_y(x)=\Re\Big[\dfrac{q}{\sqrt{\pi}}\int\limits_{0}^{\infty}
\exp\Big(-t^2+i\dfrac{x\omega_1}{t}\Big)dt\Big].
$$

\begin{center}
  \item{}\section{Заключение}
\end{center}
В настоящей работе решена вторая задача Стокса как полупространственная
граничная задача кинетической теории
с зеркально -- диффузными граничными условиями.
В основе метода лежит идея продолжить функцию распределения в сопряженное
полупространство $x<0$ и включить в кинетическое уравнение
граничное условие в виде члена типа источника.
С помощью преобразования Фурье кинетическое уравнение сводим к
характеристическому интегральному уравнению Фредгольма
второго рода, которое решаем
методом последовательных приближений.
Для этого разлагаем в ряды по степеням коэффициента диффузности
функцию распределения и массовую скорость. Подставляя эти
разложения в характеристическое уравнение и приравнивая
коэффициенты при одинаковых степенях коэффициента диффузности,
получаем счетную систему зацепленных уравнений, из которых
находим все коэффициенты искомых разложений.
В работе \cite{LY-2012_1} на примере классической задачи Крамерса
показана эффективность применяемого метода.

\begin{figure}[b]\center
\includegraphics[width=16.0cm, height=10cm]{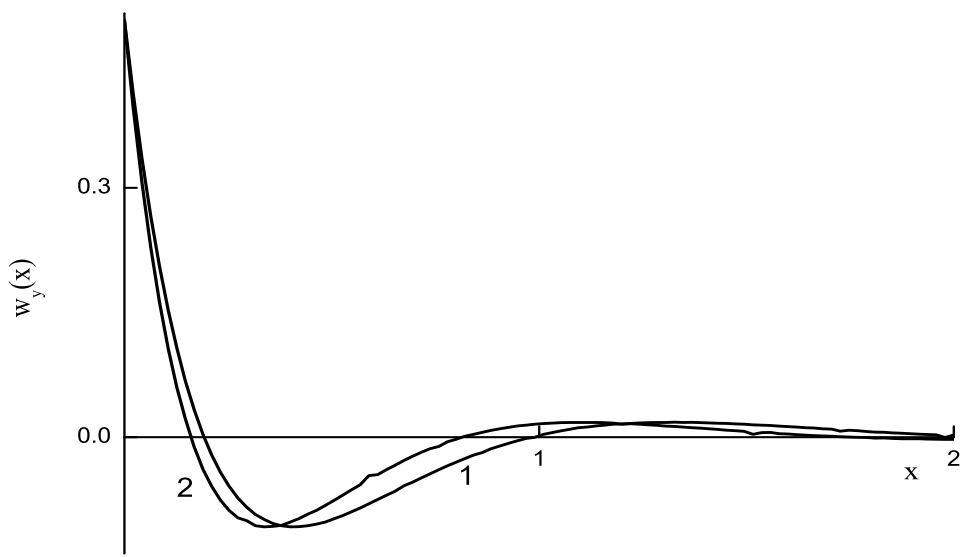}
\noindent\caption{Нормированная скорость газа в нулевом приближении,
кривые $1$ и $2$ отвечают значениям безразмерной частоты $\omega_1=5$
и $\omega_1=6$.
}\label{rateIII}
\end{figure}
\begin{figure}[h]\center
\includegraphics[width=16.0cm, height=10cm]{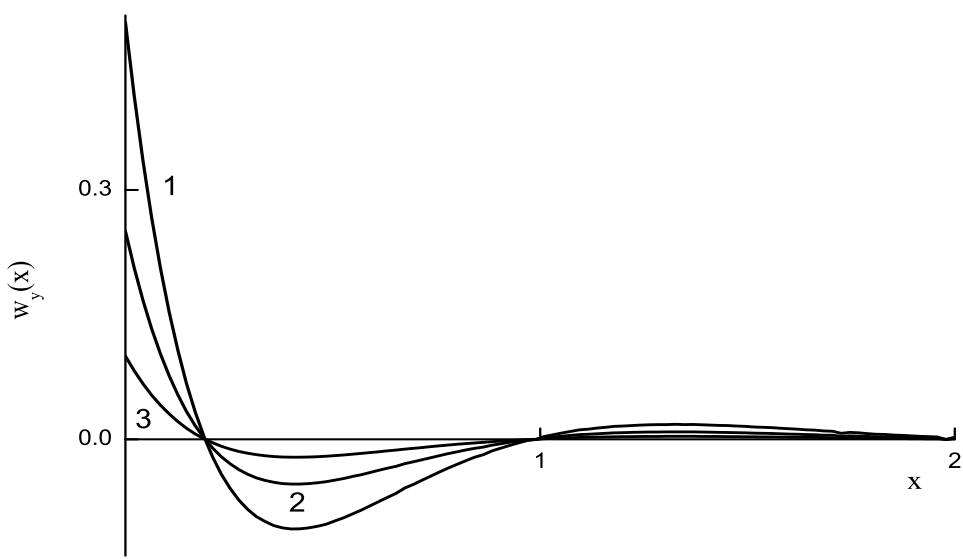}
\noindent\caption{Нормированная скорость газа в нулевом приближении,
кривые $1,2$ и $3$ отвечают значениям коэффициента аккомодации $q=1,0.5,0.2$.
}\label{rateIII}
\end{figure}
\clearpage

\end{document}